\documentclass[aps,prb,twocolumn,superscriptaddress,showpacs,floatfix]{revtex4}

\usepackage{amsmath,amssymb}
\usepackage{graphics,epsfig}

\newcommand{\varH}{{\mathcal{H}}}

\newcommand{\bfm}{{\mathbf{m}}}
\newcommand{\bfn}{{\mathbf{n}}}
\newcommand{\bfp}{{\mathbf{p}}}
\newcommand{\bfr}{{\mathbf{r}}}
\newcommand{\bfv}{{\mathbf{v}}}
\newcommand{\bfx}{{\mathbf{x}}}
\newcommand{\bfy}{{\mathbf{y}}}
\newcommand{\bfz}{{\mathbf{z}}}

\newcommand{\bfsigma}{{\boldsymbol{\sigma}}}

\newcommand{\ket}[1]{\left|#1\right\rangle}
\newcommand{\avg}[1]{\left\langle#1\right\rangle}

\newcommand{\ode}[3][]{\frac{d^{#1}{#2}}{d{#3}^{#1}}}

\newcommand{\ux}        {{\hat\bfx}}
\newcommand{\uy}        {{\hat\bfy}}
\newcommand{\uz}        {{\hat\bfz}}

\makeatletter
\newcommand{\sgn}{\mathop{\operator@font sgn}}
\makeatother

\newcommand{\eqnref}[1]{Eq.~(\ref{#1})}
\newcommand{\eqnsref}[1]{Eqs.~(\ref{#1})}
\newcommand{\figref}[1]{Fig.~\ref{#1}}
\newcommand{\Figref}[1]{Figure~\ref{#1}}
\newcommand{\secref}[1]{Sec.~\ref{#1}}
\newcommand{\Secref}[1]{Section~\ref{#1}}

\begin{document}
\title{Charge and spin density modulations in semiconductor quantum wires}
\author{Minchul Lee}
\affiliation{Department of Physics and Astronomy, University of Basel, CH-4056 Basel, Switzerland}
\affiliation{Department of Physics, Korea University, Seoul 136-701, Korea}
\author{Christoph Bruder}
\affiliation{Department of Physics and Astronomy, University of Basel, CH-4056 Basel, Switzerland}

\begin{abstract}
  We investigate static charge and spin density modulation patterns
  along a ferromagnet/semiconductor single junction quantum wire in
  the presence of spin-orbit coupling.  Coherent scattering theory is
  used to calculate the charge and spin densities in the ballistic
  regime.  The observed oscillatory behavior is explained in terms of
  the symmetry of the charge and spin distributions of eigenstates in
  the semiconductor quantum wire.  Also, we discuss the condition that
  these charge and spin density oscillations can be observed
  experimentally.
\end{abstract}

 \maketitle

\section{Introduction}

Relativistic quantum mechanics has predicted a rapid oscillatory
behavior of free electrons, called \textit{Zitterbewegung}, which
is due to the interference between the positive- and negative-energy
components in the wave packet.\cite{Sakurai67} This peculiar
oscillation happens because the velocity is not a constant of
motion even in the absence of any potential and has a fluctuating
component, while the momentum commutes with the Hamiltonian. Recently,
from the analogy between the band structure of semiconductor quantum
wells and relativistic electrons in vacuum, it has been pointed
out that electrons in semiconductors undergo the same oscillatory
motion.\cite{Schliemann04,Zitterbewegung} The essential mechanism
causing such a motion is the spin-orbit (SO) coupling in the
semiconductor. In a two-dimensional electron gas confined in a
heterostructure quantum well, two SO coupling effects are usually
taken into account: the Rashba \cite{Rashba84} and Dresselhaus
\cite{Dresselhaus55} SO coupling effects described by the following
expressions
\begin{align}
  \label{eq:SOH}
  \varH_R = \frac{\hbar k_R}{m} (\sigma_xp_y{-}\sigma_yp_x)\ \ \text{and}\ \
  \varH_D = \frac{\hbar k_D}{m} (\sigma_yp_y{-}\sigma_xp_x),
\end{align}
respectively, where $\bfsigma=(\sigma_x,\sigma_y,\sigma_z)$ are the
Pauli matrices. The strength of each SO coupling is measured in terms
of characteristic wave vectors $k_R$ and $k_D$, respectively.  The
Rashba term $\varH_R$ arises when the confining potential of the
quantum well lacks inversion symmetry, while the Dresselhaus term
$\varH_D$ is due to bulk inversion asymmetry. In the presence of
SO coupling, the velocity of the electron is not a constant of
motion:
\begin{align}
  \ode{\bfv}{t} = \frac{2\hbar(k_R^2-k_D^2)}{m^2}\, (\bfp\times\uz) \sigma_z\,.
\end{align}
The spin precession $\sigma_z(t)$ due to the SO coupling eventually
leads to a fluctuating velocity and consequently an oscillating motion of
the electron.

While the oscillatory behavior of the electron is dynamic in the
original proposal of the Zitterbewegung, it is highly plausible that
one can observe \textit{static} patterns of charge density oscillation
in some properly-structured semiconductor samples.

In our paper we investigate static charge and spin density
oscillations along a ferromagnet/semiconductor single junction quantum
wire in the ballistic limit. The charge and spin densities across the
system are calculated using coherent scattering theory. The
observed oscillatory behaviors are explained in terms of the symmetric
or antisymmetric structure of charge and spin distributions of
eigenstates in the semiconductor quantum wire. Also, we discuss the
condition that these charge and spin density oscillations can be
observed experimentally.

Our paper is organized as follows: In \secref{sec:qw} we investigate
the properties of eigenstates of semiconductor quantum wires, paying
attention to their symmetric structures in charge and spin density
distributions. The effect of SO coupling on the structures in the
presence of a confinement potential is examined in detail.
\Secref{sec:csm} is devoted to the numerical calculation of charge and
spin density modulations in the ferromagnet/semiconductor single
junction system in the presence of SO coupling. The main results
are summarized in \secref{sec:c}.

\section{Property of Eigenstates in Semiconductor Quantum Wires\label{sec:qw}}

Previous works \cite{Mireles01,Governale02} have shown that
perturbation theory cannot correctly explain the effect of moderate or
large SO coupling on energy levels and properties of eigenstates in
semiconductor quantum wires.  Instead, truncating the Hilbert space
to the lowest bands \cite{Governale02,Schliemann04} or tight-binding
models \cite{Mireles01} were successfully used to investigate the role of SO
coupling on the transport through quantum wires.
Here, we have implemented an exact numerical method
of solving the system with arbitrary strength of SO coupling. Before
introducing the method and discussing the properties of the states
obtained numerically, we briefly review the structure of the system
Hamiltonian and its symmetry in terms of transverse modes.

\subsection{Hamiltonian and Symmetry Properties}

We consider a quasi-one-dimensional system of electrons in the
presence of SO coupling:
\begin{align}
  \label{eq:H}
  \varH_{\rm S} = \frac{\bfp^2}{2m} + V(\bfr) + \varH_{\rm so},
\end{align}
where $\bfr$ and $\bfp$ are two-dimensional position and momentum
vectors and $m$ is the effective mass of the electrons in the
semiconductor. The electrons are confined in the $y$ direction by an
infinite square-well potential of width $L$,
\begin{align}
  V(\bfr) =
  \begin{cases}
    0 & (|y| < L/2)
    \\
    \infty & (|y| > L/2).
  \end{cases}
\end{align}
We assume that the SO Hamiltonian $\varH_{\rm so}$ consists of
$\varH_R$ and $\varH_D$, see \eqnref{eq:SOH}.
In some semiconductor heterostructures
(e.g., InAs quantum wells) $\varH_R$ dominates\cite{Das90a}, and in
others (e.g., GaAs quantum wells) $\varH_D$ is comparable to
$\varH_R$\cite{Lommer88a}.  A series of experiments \cite{RashbaExp}
has demonstrated that the strength of the Rashba SO coupling can be
tuned by external gate voltages. Note that our choice of the
square-well confinement eliminates the possibility of SO coupling due
to effective electric fields coming from the nonuniformness of the
confining potential. This kind of SO coupling should be considered in
case of parabolic confining potential.\cite{Moroz99}

It is instructive to rewrite the Hamiltonian in \eqnref{eq:H} in
second-quantized form to gain insight into the effect of SO coupling
on the transverse modes. To this end, we define
$c_{k_xn\mu}$/$c_{k_xn\mu}^\dag$ to be the annihilation/creation
operators of the $n$th transverse mode $\ket{k_x,n,\mu}$ with a wave
vector $k_x$ and a spin branch index $\mu=\pm$ in the \textit{absence}
of SO couplings.  It is convenient to choose the spin polarization
axis $\hat\bfn=(\cos\varphi,\sin\varphi)$ to be along the effective
magnetic field due to the SO coupling for waves propagating in the
$x$-direction such that
\begin{align}
  \label{eq:spinor}
  \ket{\mu} = \frac{1}{\sqrt2}
  \begin{bmatrix}
    \mu\, e^{-i\varphi/2}
    \\
    e^{i\varphi/2}
  \end{bmatrix}
\end{align}
with $\varphi \equiv \arg[k_D + ik_R]$.\cite{Lee04} In terms of these
operators, the Hamiltonian is expressed as
\begin{align}
  \nonumber
  \varH^{\rm 2nd}_{\rm S} & = \sum_{k_x} \left[ \sum_{n\mu} \epsilon_{n\mu}(k_x) c_{k_xn\mu}^\dag c_{k_xn\mu} \right.
  \\
  \label{eq:H2nd}
  & \qquad\qquad\left.\mbox{}
    + \sum_{nn'\mu\mu'} k_{nn'} w_{\mu\mu'} c_{k_xn\mu}^\dag c_{k_xn'\mu'} \right].
\end{align}
This expression shows that SO coupling leads to two effects. On one
hand, it lifts the spin degeneracy by
shifting the energies of the transverse modes such that
\begin{align}
  \label{eq:energyshift}
  \epsilon_{n\mu}(k_x) \equiv \frac{\hbar^2}{2m}
  \left[ (k_x - \mu k_{\rm so})^2 +
  \left(\frac{n\pi}{L}\right)^2\right] -
\Delta_{\rm so}
\end{align}
with $k_{\rm so} \equiv \sqrt{k_R^2 + k_D^2}$ and $\Delta_{\rm so}
\equiv \hbar^2 k_{\rm so}^2/2m$. On the other hand, it mixes the
transverse modes according to the amplitudes $k_{nn'} w_{\mu\mu'}$ given by
\begin{subequations}
  \label{eq:mixingamp}
  \begin{align}
    k_{nn'} & \equiv \frac{4}{L}\frac{nn'}{n^2-n^{\prime2}}
    \begin{cases}
      (-1)^{\frac{n+n'-1}{2}} & (n \ne n'\!\!\! \mod 2)
      \\
      0 & \text{otherwise}
    \end{cases}
    \\
    \label{eq:mixingampw}
    w & \equiv \frac{\hbar^2}{mk_{\rm so}}
    \begin{bmatrix}
      2i k_Rk_D & k_D^2 - k_R^2
      \\
      k_R^2 - k_D^2 & -2i k_Rk_D
    \end{bmatrix}.
  \end{align}
\end{subequations}
As shown in \eqnref{eq:mixingamp}, the SO coupling mixes transverse
modes with \textit{opposite} parities and possibly \textit{opposite}
spins. Therefore, the eigenstates of the Hamiltonian in
\eqnref{eq:H2nd} are generally not symmetric or antisymmetric.

Still, the charge and spin density pattern can be symmetric
or antisymmetric: For simplicity, we focus on the case when only one of two
SO coupling terms exists, that is, $k_Rk_D = 0$. In this case, for any
symmetric confinement potential satisfying $V(y) = V(-y)$, the
Hamiltonian commutes with the so-called ``spin parity'' operator $P_s
\equiv P_y (\hat\bfn\cdot\bfsigma)$, where $P_y$ is the inversion
operator for the $y$-component such that $P_y \psi(x,y) =
\psi(x,-y)$.\cite{Bulgakov99} The eigenstates of the system should
thus also be eigenstates of the spin parity operator, having the
eigenvalues $\pm 1$, which is evident from the fact that $P_s^2 =
1$. Denoting $\Psi_{k_xns}(\bfr)$ to be the wave function of an
eigenstate for wave vector $k_x$, subband index $n$, and quantum
number of spin parity $s=\pm1$, one can show that the spin density
components $S_{k_xns}^{\hat\bfm}(\bfr) \equiv \Psi_{k_xns}^\dag(\bfr)
(\hat\bfm\cdot\bfsigma) \Psi_{k_xns}(\bfr)$ for
$\hat\bfm\perp\hat\bfn$ should be antisymmetric with respect to
$y=0$,\cite{Governale02,Debald04} while the charge density and the
spin component $S_{k_xns}^{\hat\bfm}$ for $\hat\bfm = \hat\bfn$ are
symmetric. For example, $S_{k_xns}^\uz(y) = \Psi_{k_xns}^\dag(-y)
(\hat\bfn\cdot\bfsigma) \sigma_z (\hat\bfn\cdot\bfsigma)
\Psi_{k_xns}(-y) = - S_{k_xns}^\uz(-y)$. This (anti)symmetric
properties for $\hat\bfm=\uz$ persist even in case $k_Rk_D\ne0$, when
there is no simple operator like $P_s$ commuting with the
Hamiltonian. However, as $|k_R|$ and $|k_D|$ are tuned to come close
to each other, the antisymmetric components decrease in magnitude and
even vanish when $|k_R|=|k_D|$ since the off-diagonal terms in the
matrix $w$ that couple modes with opposite spins
vanish. Interestingly, for $|k_R|=|k_D|$, the spin operator
$\hat\bfn\cdot\bfsigma$ itself commutes with the Hamiltonian and the
SO coupling couples only the transverse modes with the \textit{same}
spin
(see \eqnref{eq:mixingampw}) so that a common spin quantization axis
can be defined for all the eigenstates.\cite{Schliemann03}

The symmetry properties discussed up to now are not restricted to the
square-well potential case. One can show that the same
reasoning works in the case of a harmonic potential.\cite{Debald04} All
that one should do is to redefine the amplitude $k_{nn'}$ according to
the confinement potential. For the symmetric potential satisfying
$V(-y) = V(y)$, the amplitude $k_{nn'}$ has nonzero values only when
$n$ and $n'$ have opposite parities and reproduces the same symmetry
properties as discussed above. Hence, the choice of confinement
potential does not lead to any qualitatively different
effect. Therefore, we focus on the square-well
potential since it is convenient for numerical calculations.

\subsection{Numerical Calculation of Eigenstates\label{sec:eigenstate}}

It is possible to obtain the eigenstates numerically by diagonalizing
the Hamiltonian \eqnref{eq:H2nd} in a truncated basis of transverse
modes. Since we need eigenstates at a given energy, not a given wave vector,
we have adopted another numerical
method to obtain the eigenstates: First, prepare eigenstates at a
given energy in the \textit{absence} of the confinement
potential\cite{Lee04} and then numerically find a set of wave vectors
$k_x$ such that the linear superposition of four eigenstates sharing
the same wave vector $k_x$ satisfies free boundary conditions at the
infinite walls, that is, vanishes at $y=\pm L/2$. This method is
superior to the diagonalization of the Hamiltonian $\varH^{\rm
2nd}_{\rm S}$ in two aspects; one can obtain numerically exact
eigenstates in this way,
and an infinite set of
evanescent waves, that is, eigenstates with complex $k_x$ at the
targeted energy can be found systematically. Evanescent waves
turned out to be important in transport through (semi-)finite
samples.\cite{Lee04}

\begin{figure}[!t]
  \centering
  \includegraphics[width=8.5cm]{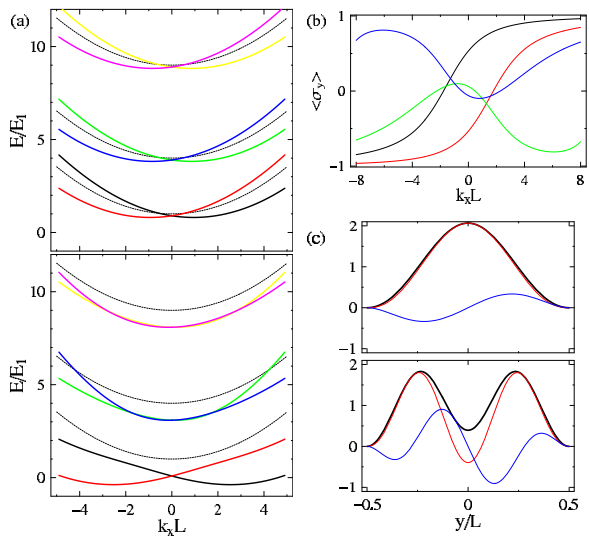}
  \caption{(Color online) (a) Spectra of a quantum wire confined by a
    square-well potential in the presence of Rashba spin-orbit (SO)
    coupling: $k_RL = 1$ (upper) and 3 (lower). The dotted lines
    correspond to the modes in the absence of SO coupling. The energy
    is scaled by $E_1=\frac{\hbar^2\pi^2}{2mL^2}$, the ground-state
    energy at $k_x=0$ in the absence of SO coupling. (b) Spin
    expectation value $\avg{\sigma_y}$ for four low-lying states at
    $k_RL=3$. Same colors as in (a) are used to identify the
    levels. (c) Charge (black), spin-$y$ (red), and spin-$z$ (blue)
    density profiles along the $y$-direction for the lowest subband
    ($n=1$) with $s=+1$ (upper) and $-1$ (lower) at $k_RL = 3$ and
    $E/E_1 \approx 1.74^2$.}
  \label{fig:levels}
\end{figure}

\begin{figure}[!tb]
  \centering
  \includegraphics[width=8.5cm]{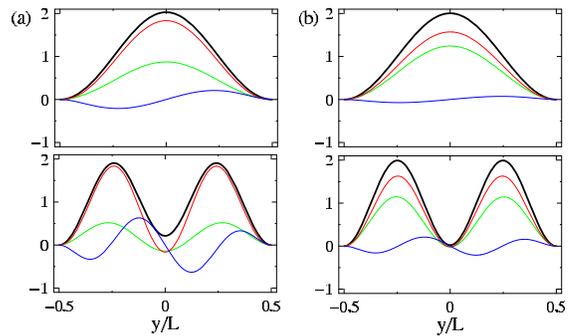}
  \caption{(Color online) Charge (black), spin-$x$ (green), spin-$y$
    (red), and spin-$z$ (blue) density profiles along the
    $y$-direction for the lowest subband ($n=1$) with $s=+1$ (upper)
    and $-1$ (lower) at $(k_RL,k_DL) =$ (a) (3,1) and (b) $(3,2)$ and
    at energy $E/E_1 \approx 1.74^2$.}
  \label{fig:profiles}
\end{figure}

First, we consider the case where only Rashba SO coupling is
present ($k_D=0$).  \Figref{fig:levels} shows the spectra of this
system and the properties of the eigenstates for weak and strong SO
coupling.
Small SO coupling ($k_RL\lesssim1$) splits the spin degeneracy
according to \eqnref{eq:energyshift} (see the upper figure in
\figref{fig:levels}(a)) and each spin branch $s=\pm1$ keeps
its spin polarization direction approximately such that
$\avg{(\hat\bfn\cdot\bfsigma)}_{k_xns} = \avg{\sigma_y}_{k_xns}
\approx s$ as long as $k_x$ is not so large ($k_x<(2k_RL^2)^{-1}$).
Due to the antisymmetry of the spin distribution, the spin
expectation values $\avg{\sigma_x}$ and $\avg{\sigma_z}$ strictly
vanish irrespective of the SO coupling strength.  However, the spin
density component $S_{k_xns}^{\uz}(\bfr)$ is finite and has opposite
sign for different branches: $S_{k_xns}^{\uz}(\bfr) \approx
-S_{k_xn-s}^{\uz}(\bfr)$.
As $k_RL$ increases, on the other hand, the mixing of the transverse
modes with opposite spins becomes more apparent such that level
crossings happen even in the same subband for $n>1$ (see the lower
figure in \figref{fig:levels}(a)) and the spin polarization is changed
and even reversed compared to the case of weak SO coupling.
\Figref{fig:levels}(b) shows that at strong SO coupling $(k_RL=3)$ and
large $k_x (\gtrsim (2k_RL^2)^{-1})$ the spins of two branches in the
lowest subband $(n=1)$ are polarized parallel to each other:
$\avg{\sigma_y}_{n=1,k_xs} \approx \sgn(k_x)$ for both
$s=\pm1$.\cite{Governale02} This is due to mixing with higher-subband
states having opposite spin, which manifests itself well in
\figref{fig:levels}(c) where the charge density profile of the state
$s=-1$ of the lowest subband exposes the contribution from the higher
subband $n=2$. The charge and spin density profiles in
\figref{fig:levels}(c) also confirm the (anti)symmetric structure
proven based on the symmetry of the system.

Dresselhaus SO coupling $\varH_D$, related to $\varH_R$ by a
unitary transformation $U = (\sigma_x+\sigma_y)/\sqrt2$, gives rise to
an energy spectrum and charge and spin density profiles identical to
those obtained in the Rashba SO coupling case except for the rotation of
spin axes such that $x\to y$, $y\to x$, and $z\to -z$ and the
substitution of $k_R$ by $k_D$.

In the presence of both Rashba and Dresselhaus SO coupling terms, the
spin polarization of the eigenstates is not along the $x$- or $y$-axes any
longer, since there are nonzero spin expectation values $\avg{\sigma_x}$ and
$\avg{\sigma_y}$, and its direction depends on the ratio
$k_R/k_D$. Nevertheless, as
shown in \figref{fig:profiles}, the spin density
profiles are found to have (anti)symmetric structures similar to the
previous cases: $S^{\ux/\uy}(y) = S^{\ux/\uy}(-y)$ and $S^\uz(y) =
-S^\uz(-y)$. Note that both spin-$x$ and $y$ components are
symmetric. The antisymmetric component $S^\uz(y)$ is observed to
decrease in magnitude as $k_D$ approaches $k_R$, and eventually
vanishing when $k_R=k_D$. This behavior, consistent with the argument
in terms of coupling matrix elements of $w$ in \eqnref{eq:mixingampw},
indicates that whereas Rashba and Dresselhaus SO coupling
can induce an antisymmetric distribution of $S^\uz(y)$ separately, the
presence of both of them will lead to a reduction in
$S^\uz(y)$.

In summary, SO interaction in the presence of a symmetric confining
potential can result in antisymmetric structures in the spin distributions
of individual eigenstates, especially in the spin-$z$ component. This
phenomenon manifests itself strongly when either Rashba or Dresselhaus
SO coupling is present and vanishes as both coupling strengths become
equal to each other.

\section{Charge and Spin Modulations in the Presence of Spin-orbit Coupling\label{sec:csm}}

In this section we investigate the charge and spin density modulations
in a ferromagnet/semiconductor single junction quantum wire. Static
charge and spin density oscillations are induced by injecting a
spin-polarized current from ferromagnet to semiconductor. We make use
of the coherent scattering theory to calculate the charge and spin
density modulations in the ballistic regime and uncover the fact that
the (anti)symmetric structure of eigenstates in semiconductor quantum
wires studied in the previous section is deeply related to such
density oscillations.

\subsection{Scattering Theory}

The single junction system consists of the ferromagnet (F; $x{<}0$)
and the semiconductor (S; $x{>}0$) regions confined by the square-well
potential of width $L$. While the electrons in the semiconductor
region are governed by \eqnref{eq:H}, the Hamiltonian for the
ferromagnet region is
\begin{align}
  \varH_{\rm F} = \frac{\bfp^2}{2m} + V(\bfr) - (h/2)(\sigma_z - 1).
\end{align}
The spin-splitting energy $h$ is assumed to be very large compared to
other energy scales such that only the majority spin state (spin up
state) is available at the energy of interest.  Here we have assumed
that both regions have identical effective masses and lower band edges,
because the impedance mismatch due to a difference in them does not
affect the density oscillation in a qualitative way. In order to
set up the scattering theory one should construct a set of
eigenstates in both regions at a given energy $E$. Denote
$\Psi_{\epsilon m}^{F/S}(\bfr) = e^{i\epsilon k_{mx}^{F/S}x}
\Phi_m^{F/S}(y)$ to be eigenstates in the ferromagnet/semiconductor
regions propagating in the $x$ $(\epsilon=+)$ or $-x$ $(\epsilon=-)$
direction at energy $E$, where $m=(n,s)$ is a composite index with
subband index $n$ and spin index $s$ and $\Phi_m(y)$ is a spinor for
the state. Note that the wave vector $k_{mx}$ can be complex-valued
in the case of evanescent waves.

The boundary conditions at the interface $(x=0)$ are then specified by
the following two equations:
\begin{subequations}
  \label{eq:bc}
  \begin{align}
    \label{eq:bca}
    \Psi^F(\bfr)|_{x=0} & = \Psi^S(\bfr)|_{x=0}
    \\
    \label{eq:bcb}
    v_x^F \Psi^F(\bfr)|_{x=0} & = v_x^S \Psi^S(\bfr)|_{x=0}
  \end{align}
\end{subequations}
for all $y \in [-L/2,L/2]$. Here the wave functions in both regions
can be expanded in terms of the eigenstates as follows: $\Psi^F(\bfr)
= \sum_m c^{(i)}_m \Psi_{+m}^F(\bfr) + \sum_m c^{(r)}_m
\Psi_{-m}^F(\bfr)$ and $\Psi^S(\bfr) = \sum_m c^{(t)}_m
\Psi_{+m}^S(\bfr)$ with complex coefficients $c^{(i,r,t)}_m$. The
velocity operators, $v_x^F = p_x/m$ and $v_x^S = (p_x -
\hbar(k_R\sigma_y+k_D\sigma_x))/m$ differ from each other due to the
presence of SO coupling in the semiconductor.

The SO coupling deforms the transverse modes such that they do not
match in the ferromagnet and semiconductor regions.  It is thus
impossible to reduce the scattering problem to a one-dimensional one.
Instead we set up an infinite number of coupled linear equations for
the coefficients $c^{(i,r,t)}$ by multiplying $\Phi_m^F(y)$ on both
sides of \eqnsref{eq:bca} and (\ref{eq:bcb}) and integrating them over
$y$. The coupled linear equations are then solved numerically for the
reflection/transmission coefficients $c^{(r,t)}_m$ at given incident
coefficients $c^{(i)}_m$. In our study we focus on the injection of
the electron through the lowest transverse mode such that
$c^{(i)}_{(n=1,s=+)} = 1$ and the other incident coefficients are
zero.  Since the contribution of evanescent waves in high subbands is
very small in this case, the coupled equations can be truncated to be
of finite dimensions with negligible errors. By using the coefficients
obtained in this way, we calculate and investigate the charge and spin
density profiles, $\rho(\bfr) \equiv |\Psi^{F/S}(\bfr)|^2$ and
$S^{\hat\bfm}(\bfr) \equiv
\Psi^{F/S\dag}(\bfr)(\hat\bfm\cdot\bfsigma)\Psi^{F/S}(\bfr)$ in the
next sections.

\subsection{Charge Oscillation due to Spin-orbit Coupling}

\begin{figure}[!tb]
  \centering
  \begin{minipage}{7cm}
    \begin{flushright}
      \includegraphics[width=0.879\textwidth,bb = 148 65 489.518127 137.400003,clip=]{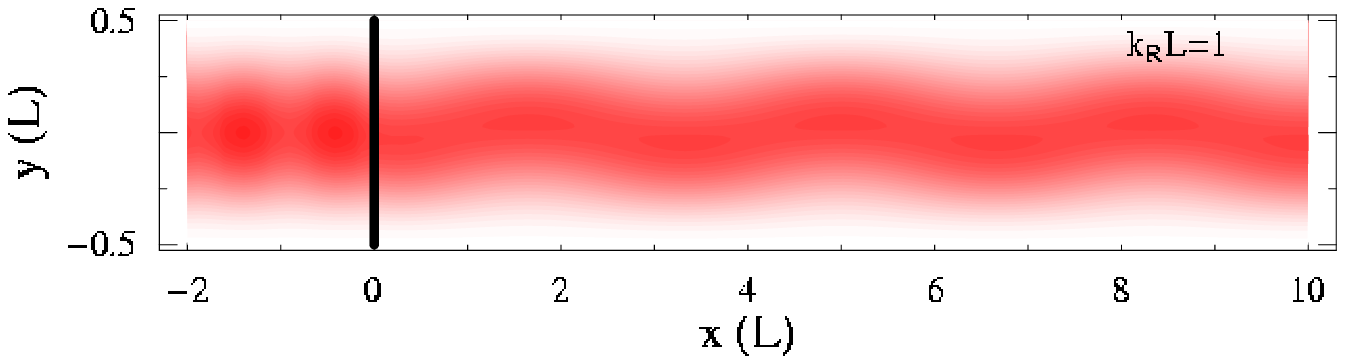}
      \includegraphics[width=0.879\textwidth,bb = 148 65 489.518127 137.400003,clip=]{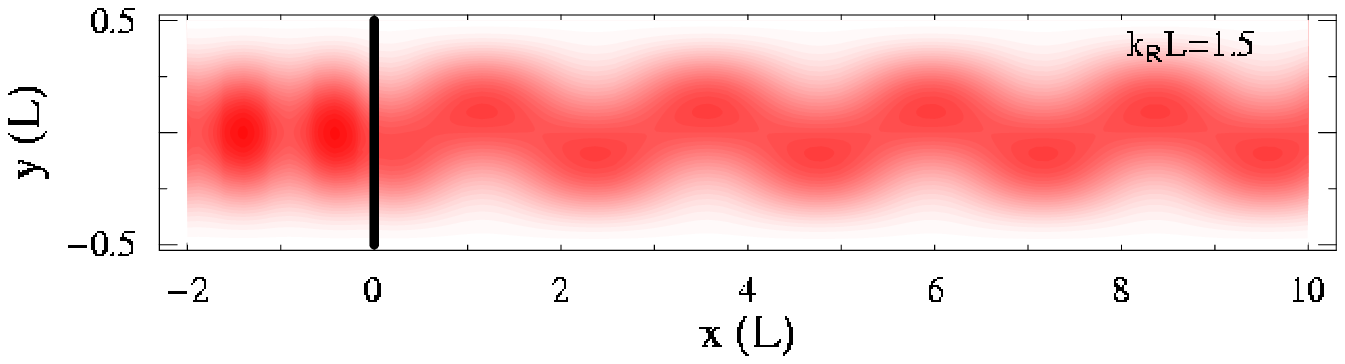}
      \includegraphics[width=0.879\textwidth,bb = 148 65 489.518127 137.400003,clip=]{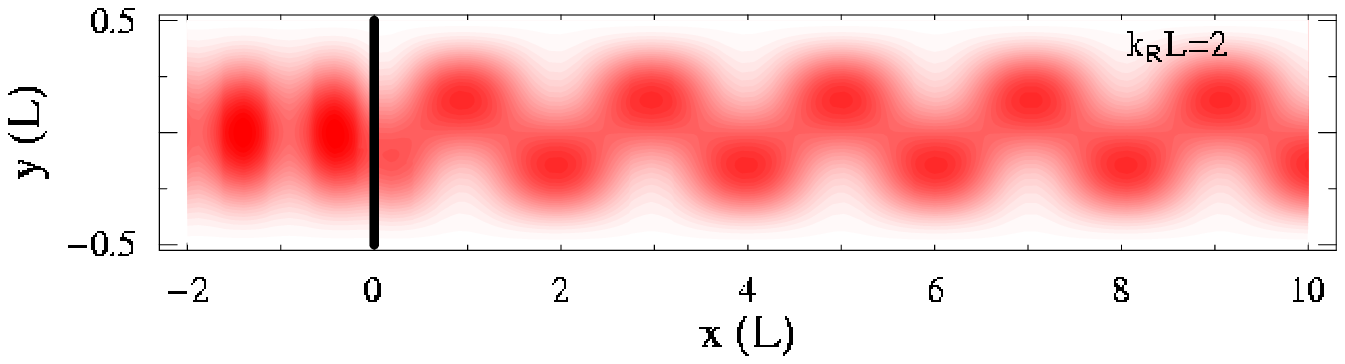}
      \includegraphics[width=0.879\textwidth,bb = 148 65 489.518127 137.400003,clip=]{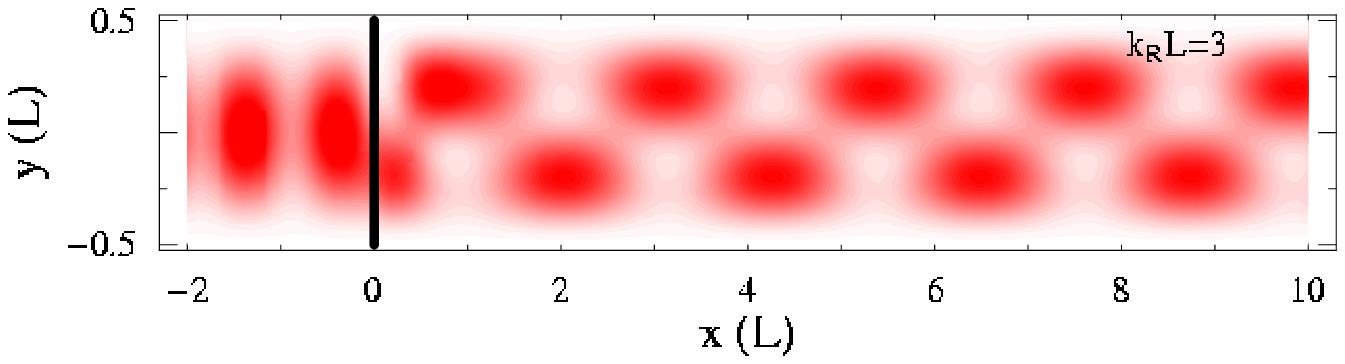}
      \includegraphics[width=\textwidth,bb = 101 34 489.518127 137.400003,clip=]{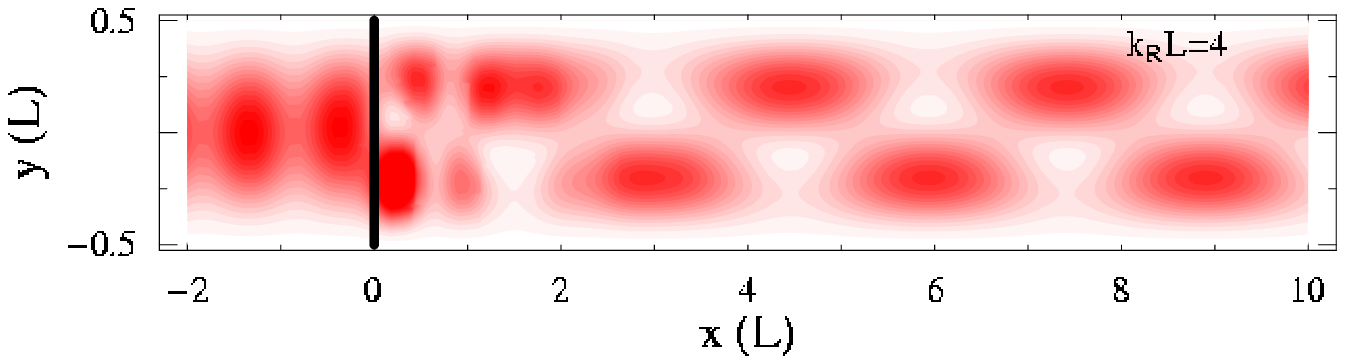}
    \end{flushright}
  \end{minipage}
  \begin{minipage}[c]{0.7cm}
    \includegraphics[width=\textwidth,bb = 88 83 138 222,clip=]{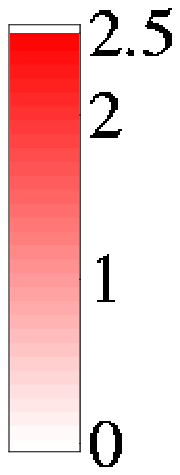}
  \end{minipage}
  \caption{(Color online) Charge density profiles along the quantum
    wire in the neighborhood of the ferromagnet/semiconductor
    interface. A spin-polarized current is injected in the
    lowest transverse mode from the ferromagnet $(x<0)$ to the
    semiconductor $(x>0)$. The distances are scaled by $L$. Here we
    have set $E/E_1 \approx 1.42^2$ and varied the strength of the
    Rashba SO coupling from $k_RL = 1$ to 4. Irregular patterns near
    the interface are caused by evanescent waves whose contributions
    increase with $k_R$.}
  \label{fig:chargeosc}
\end{figure}

First, we investigate the charge density oscillation due to Rashba
SO coupling and focus on the energy levels for which only two propagating
states exist in the semiconductor region. \Figref{fig:chargeosc} shows
the charge density profiles near the ferromagnet/semiconductor
junction for various strengths of the Rashba SO coupling and clearly
disclose the static oscillatory patterns perpendicular to the
propagating direction. The patterns oscillate around the center line
of the wire and the oscillation amplitude increases with the coupling
strength $k_RL$.  For large SO coupling $(k_RL>1.5)$ the electron
does not stay any longer near the center line, forming high-density
islands off-center. This oscillation is related to the fact that
two propagating waves are eigenstates of the spin parity operator with
opposite eigenvalues $\pm 1$. From this property one can prove that
the charge density $\rho(\bfr)$ can be divided into symmetric and
antisymmetric parts with respect to $y=0$:
\begin{align}
  \rho(x,y) = \rho_s(y) + \rho_a(x,y).
\end{align}
While the symmetric component $\rho_s(y) = \rho_s(-y)$ is uniform
along the wire, the antisymmetric part $\rho_a(x,y) = -\rho_a(x,-y)$
oscillates along the $x$-direction with period $\lambda = 2\pi/\Delta
k$ and generates the density modulations shown in \figref{fig:chargeosc}.
Here, $\Delta k$ is the wave vector difference of two propagating modes.

\begin{figure}[!tb]
  \centering
  \includegraphics[width=4cm]{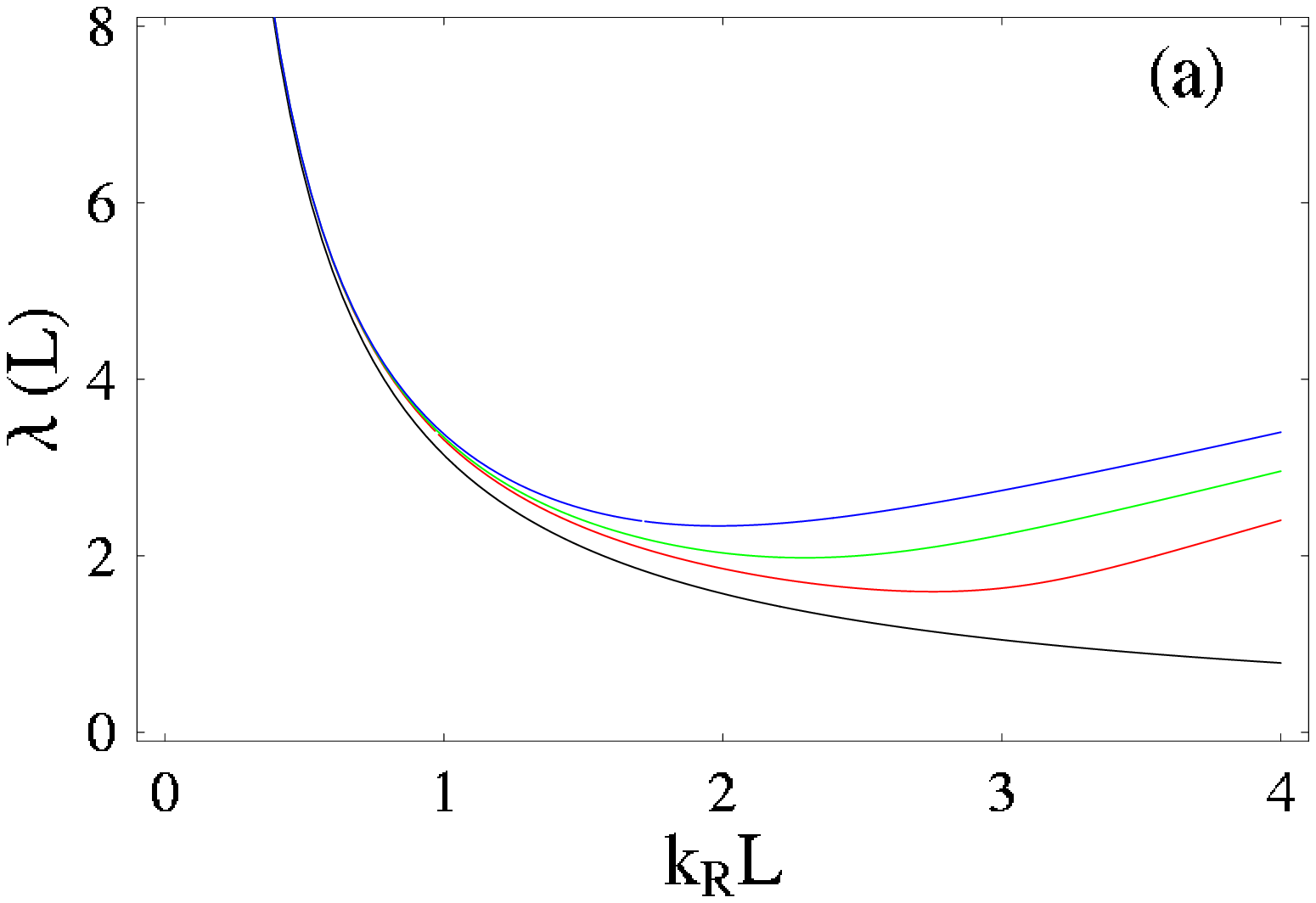}
  \includegraphics[width=4cm]{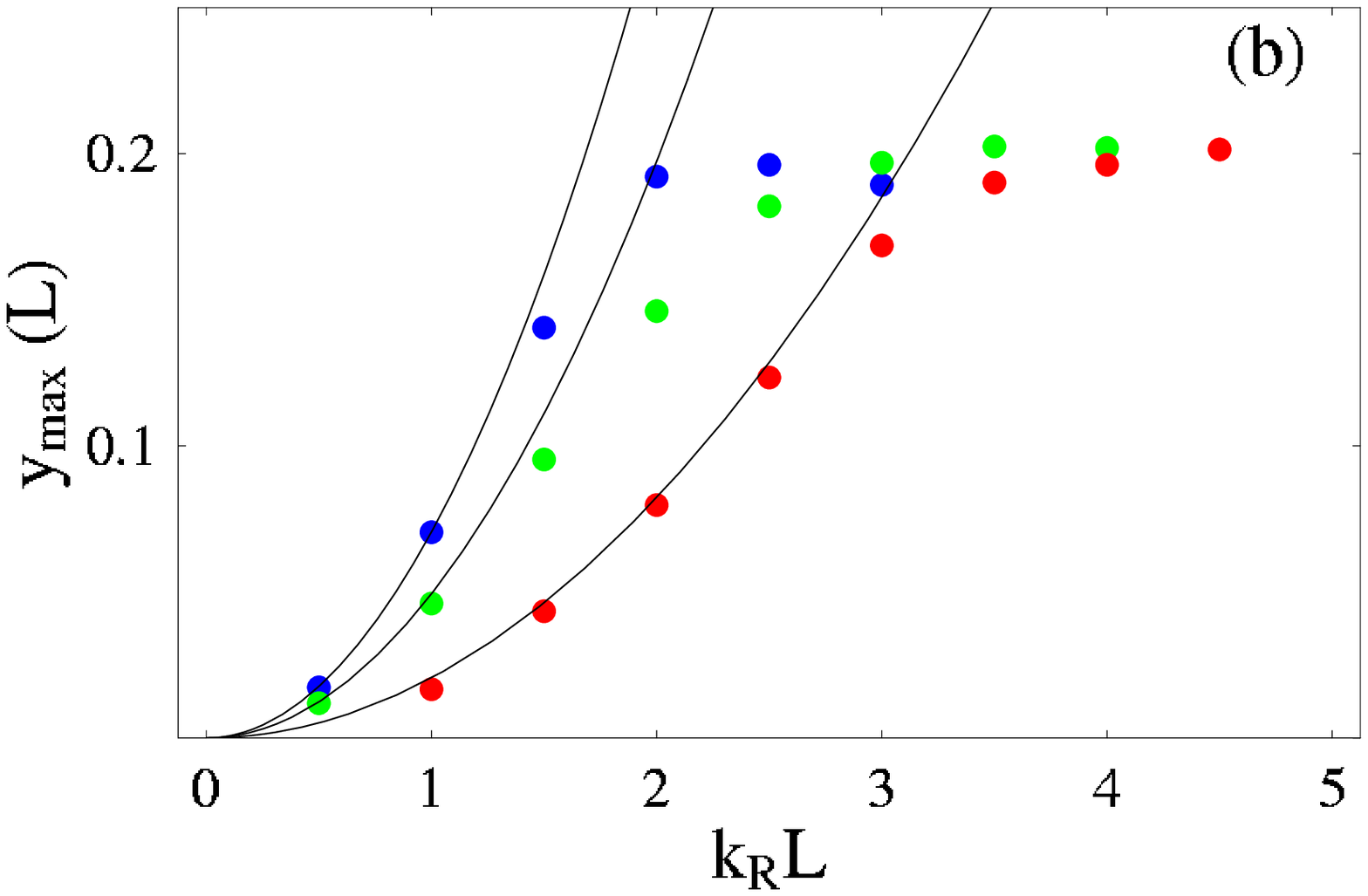}
  \caption{(Color online) (a) Period and (b) amplitude of the charge
    oscillation as functions of Rashba SO coupling strength $k_RL$ for
    various energies $E/E_1 \approx$ $1.01^2$ (red), $1.42^2$ (green),
    and $1.74^2$ (blue).  The black solid lines are obtained from
    perturbation theory such that (a) the line goes like $\lambda =
    \pi/k_R$ and (b) they show a parabolic dependence on $k_R$.}
  \label{fig:periodamp}
\end{figure}

The oscillation period $\lambda$ depends on both the SO coupling
strength and the energy of the incident electron, as shown in
\figref{fig:periodamp}(a). At weak SO couplings the period is
determined by the characteristic wave vector $k_R$ such that it
decreases like $\lambda \approx \pi/k_R$ regardless of the energy.  As
$k_R$ increases further, on the other hand, the period increases with
$k_R$, having a minimum. In addition, the period gets longer at higher
energies. The non-monotonic behavior of the period is another
manifestation of mixing between transverse modes due to the SO
coupling. \Figref{fig:periodamp}(b) shows the dependence of the
oscillation amplitude $y_{\rm max}$ on the SO coupling strength and
the energy. Here the oscillation amplitude is given by the distance
between the center line $(y=0)$ and the points where the density
reaches its maximum. The parabolic dependence of $y_{\rm max}$ on
$k_R$ at small SO couplings indicates that the charge density
oscillation is a second-order effect of the SO coupling. As $k_R$ is
increased, the amplitude does not increase any longer and instead
saturates to $y_{\rm max} \approx L/5$ as long as the condition on the
number of propagating waves is not violated.

\begin{figure}[!t]
  \centering
  \begin{minipage}{8cm}
    \centering
    \includegraphics[width=\textwidth,bb = 101 34 489.518127 137.400003,clip=]{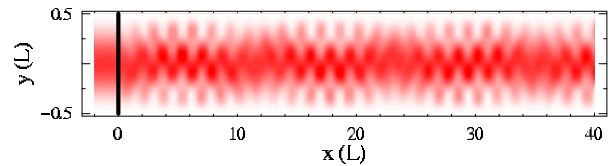}
  \end{minipage}
  \caption{(Color online) Charge density profiles at $k_RL = 2$ and
    $E/E_1 \approx 6.68^2$. The density scale is the same as in
    \figref{fig:chargeosc}.}
  \label{fig:chargeosche}
\end{figure}

At higher energies more than two propagating waves are involved in the
scattering process in the semiconductor region and the charge density
oscillation pattern due to their interference cannot be specified by
one period.  Instead, some beating patterns appear as shown in
\figref{fig:chargeosche}. While the SO coupling still leads to
oscillatory motion of the charge, oscillations of short and long
periods are mixed and the pattern becomes complicated.

Our observation is related to the prediction of Schliemann and
Loss\cite{Schliemann04} about the oscillatory behavior of free
electrons in the presence of Rashba SO coupling: the center of an
electron wave packet oscillates in the direction perpendicular to its
group velocity $\bfv$ with the frequency $2k_R|\bfv|$, giving rise to
the oscillation period $\pi/k_R$ along the propagating
direction. Controlled steady injection of wave packets with the same
$\bfv$ can produce the static density modulations we have
observed. They also studied the dynamic oscillation of an electron
with a given momentum $k_x$ in quantum wires, where the electron is
prepared initially in the lowest-lying spin-polarized states.  In this
case, since the initial state is not an energy eigenstate, dynamic
beating patterns due to the interference of waves with different
energies are observed. On the other hand, our static density
oscillation patterns are due to the interference of the energy
eigenstates with a given energy.  For that reason, such beating
patterns and the resonance phenomena, leading to the sharp peak in the
oscillation amplitude, are absent in our observation.

\begin{figure}[!t]
  \centering
  \includegraphics[width=5cm]{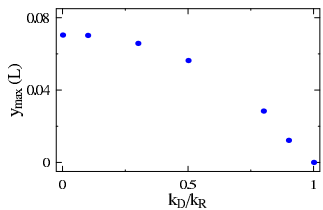}
  \caption{(Color online) Charge oscillation amplitude as a function
    of the ratio $k_D/k_R$ for $k_RL=1$ and $E/E_1 \approx 1.74^2$.}
  \label{fig:dresselhausamp}
\end{figure}

We now include the effect of the Dresselhaus SO coupling on the charge
density modulation. The unitary relation between two SO terms
$\varH_R$ and $\varH_D$ discussed in \secref{sec:qw} guarantees that
the Dresselhaus SO coupling results in exactly the same charge density
oscillation as the Rashba SO coupling.  In the presence of both
coupling terms, however, the oscillation vanishes as the strengths of
two terms become equal to each other.  \Figref{fig:dresselhausamp}
shows that the amplitude of the oscillation decreases with $k_D/k_R$
for finite $k_R$ and the oscillation disappears when $k_R=k_D$. This
tendency is quite similar to the dependence of the antisymmetric
distribution of $S^\uz(y)$ on the SO coupling. It also supports the
close relation between the charge density oscillation and the
antisymmetric structure of eigenstates in semiconductor quantum wires.

Before closing this section we address the experimental observability
of the charge density modulations. Since the oscillation patterns
depend on the energy of the incident electron, it should be checked
whether the superposition of charge density patterns from different
energies can weaken the overall density oscillation.
\Figref{fig:periodamp}(a) shows that for large SO coupling
$(k_RL\gtrsim1.5)$ the period strongly depends on the
energy. Therefore, in the case of a large voltage drop across the
junction the injection of electrons with a wide range of energy will
smear the density modulation by superposing density patterns
oscillating with different periods.  For small SO coupling, on the
other hand, the phase as well as the oscillation period pattern are
weakly dependent on the energy, leading to the superposition of
commensurate density patterns. In addition, the states below the fermi
level are irrelevant to the charge density oscillation because the
levels are filled incoherently such that their charge distributions
are always symmetric with respect to $y=0$. Hence the condition for
the observation of charge density oscillation is that either the SO
coupling strength or the applied voltage across the junction should be
small enough.  Under this condition, charge oscillations can be detected
via high-resolution scanning probe microscopy techniques\cite{STM} by
imaging the charge density directly or by measuring the change of
conductance along the quantum wire as a function of the position of a
tip that can deplete charges under it.\cite{Schliemann04}

\subsection{Spin Oscillation due to Spin-orbit Coupling}

\begin{figure}[!tb]
  \centering
  \begin{minipage}{7cm}
    \begin{flushright}
      \includegraphics[width=0.879\textwidth,bb = 148 65 489.518127 137.400003,clip=]{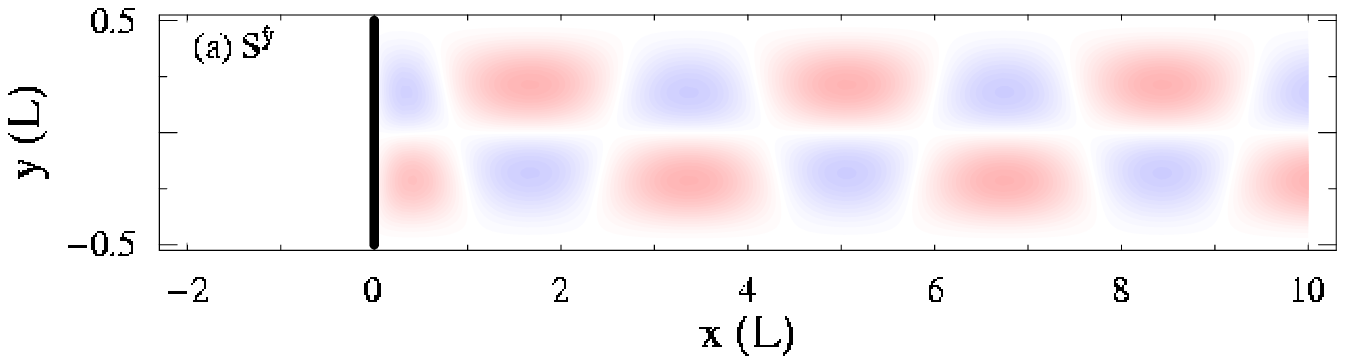}
      \includegraphics[width=0.879\textwidth,bb = 148 65 489.518127 137.400003,clip=]{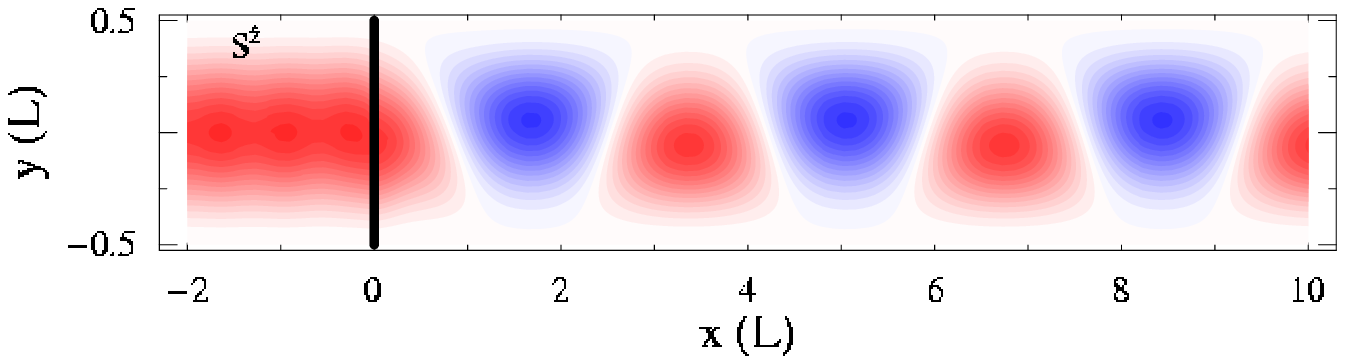}
      \includegraphics[width=0.879\textwidth,bb = 148 65 489.518127 137.400003,clip=]{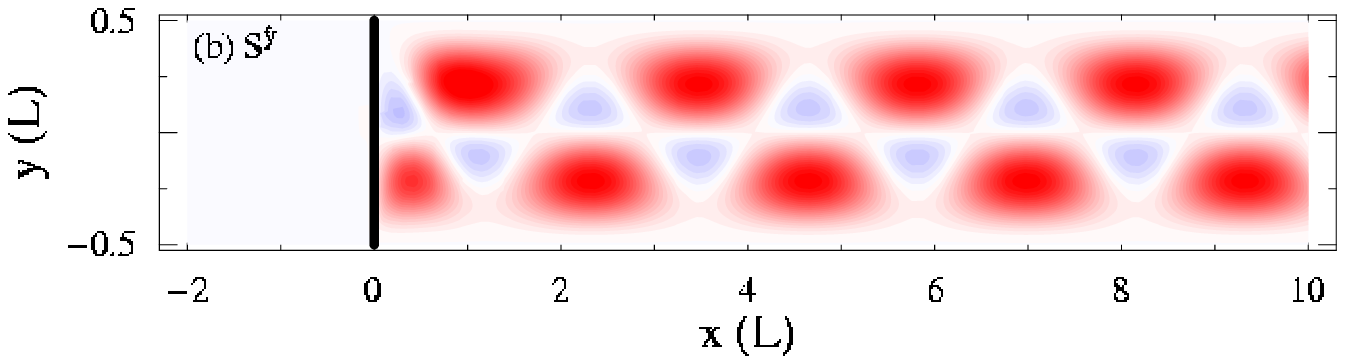}
      \includegraphics[width=\textwidth,bb = 101 34 489.518127 137.400003,clip=]{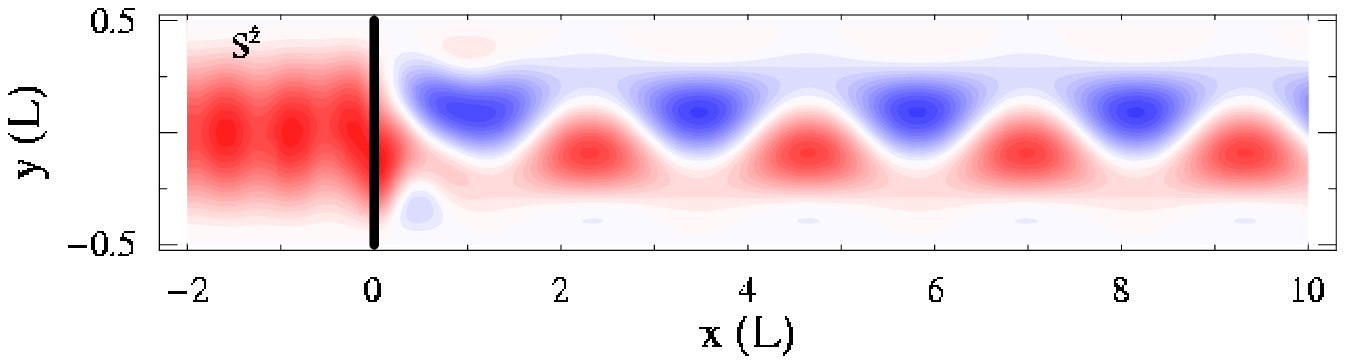}
    \end{flushright}
  \end{minipage}
  \begin{minipage}[c]{0.7cm}
    \includegraphics[width=\textwidth,bb = 88 126 153 270,clip=]{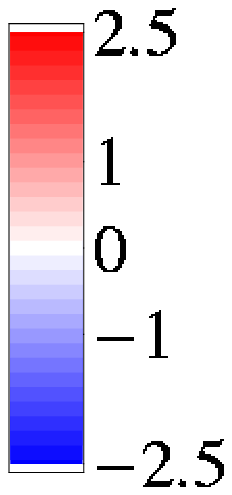}
  \end{minipage}
  \caption{(Color online) Spin density distributions, $S^\uy(\bfr)$ (upper) and $S^\uz(\bfr)$ (lower) for $k_RL = $
    (a) 1 and (b) 2 in the same configuration as used in \figref{fig:chargeosc}. Here we have set $E/E_1 \approx
    1.74^2$. Note that $S^\ux(\bfr) = 0$.}
  \label{fig:spinoscr}
\end{figure}

In the presence of SO coupling the electron spin precesses in the
effective magnetic field that depends on the momentum. Spin density
patterns thus reflect such spin precession as well as the charge
density distribution discussed in the previous
section. \Figref{fig:spinoscr} shows spin density profiles along the
quantum wire if only Rashba SO coupling is present. The spin
precessing patterns in $S^\uz(\bfr)$ indicate that the effective
magnetic field is approximately perpendicular to the electron's moving
direction. The spin distribution $S^\uz(\bfr)$ is not symmetric with
respect to $y=0$ in the semiconductor region and this behavior gets
stronger as the SO coupling increases. It is because the spin pattern
follows the charge density oscillation that is also asymmetric. While
$S^\ux(\bfr) = 0$ for all $\bfr$, $S^\uy(\bfr)$ is nonvanishing inside
the semiconductor, increasing in magnitude with the SO coupling
strength. Unlike $S^\uz(\bfr)$, $S^\uy(\bfr)$ get biased to positive
values as $k_R$ is increased, indicating a spin polarization along the
$y$-direction. This is a direct consequence of the spin polarization
of the lowest modes at large $k_x$ and $k_R$ (see
\secref{sec:eigenstate}).

\begin{figure}[!t]
  \centering
  \begin{minipage}{7cm}
    \begin{flushright}
      \includegraphics[width=0.879\textwidth,bb = 148 65 489.518127 137.400003,clip=]{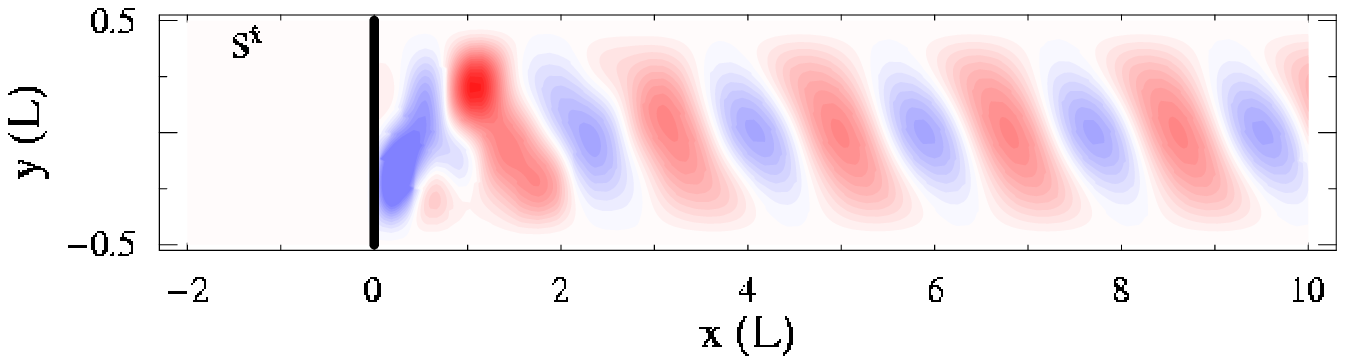}
      \includegraphics[width=0.879\textwidth,bb = 148 65 489.518127 137.400003,clip=]{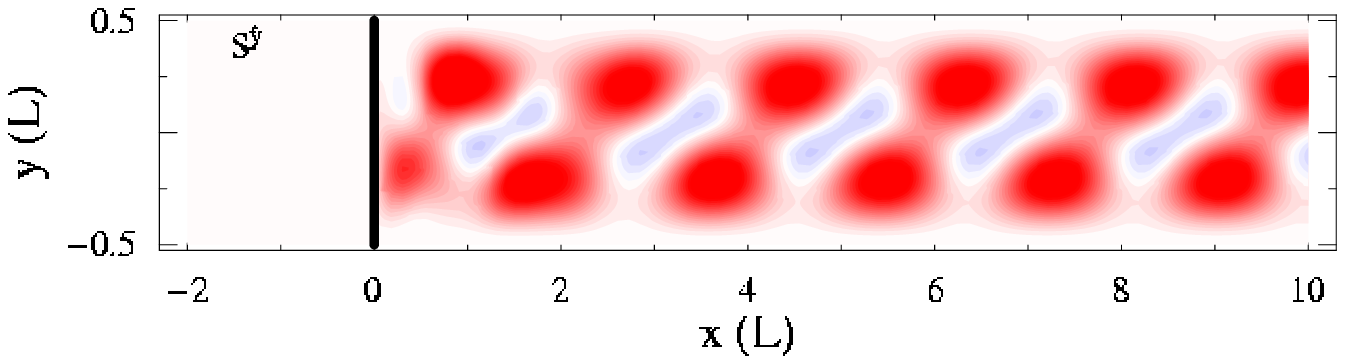}
      \includegraphics[width=\textwidth,bb = 101 34 489.518127 137.400003,clip=]{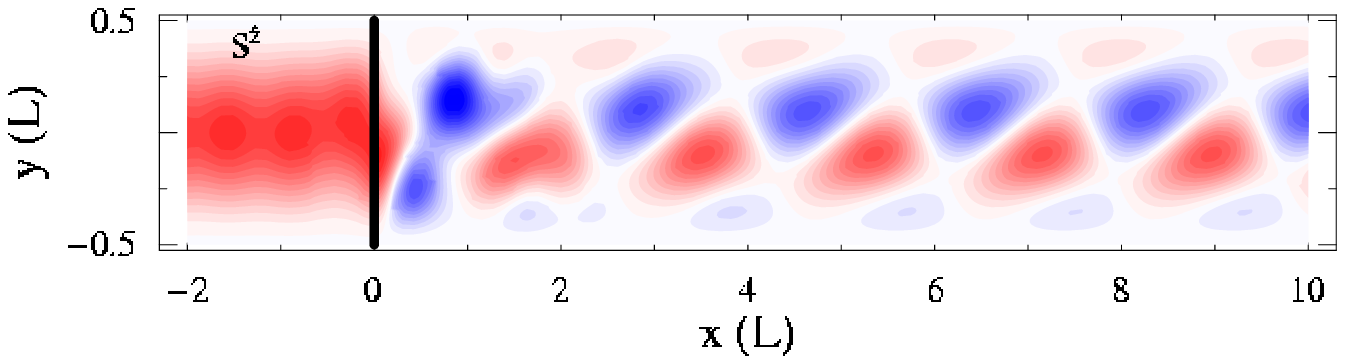}
    \end{flushright}
  \end{minipage}
  \caption{(Color online) Spin density distributions,
    $S^\ux(\bfr)$(upper), $S^\uy(\bfr)$ (middle), and $S^\uz(\bfr)$
    (lower) in the same configuration as used in
    \figref{fig:chargeosc}. Same parameters and color scaling as in
    \figref{fig:spinoscr} are used except that $k_RL = 2$ and $k_DL =
    1$.}
  \label{fig:spinoscdr}
\end{figure}

The Dresselhaus SO coupling also gives rise to similar spin profiles;
the spin axes $x$ and $y$ are exchanged, and the spin distributions
are mirror-reflected with respect to $y=0$. In the presence of both
coupling effects, on the other hand, the spin density patterns, with
all the components nonvanishing, become distorted as shown in
\figref{fig:spinoscdr}.  Slanted patterns appear due to the
absence of any symmetry with respect to $y=0$. As long as $k_R\ne
k_D$, the spin is preferentially polarized along either $x$- (when
$k_D > k_R$) or $y$- (when $k_R < k_D$) direction. For $k_R=k_D$
an unbiased spin precession takes place for all the spin components.

Under the same condition found for the detection of charge density
modulations, the spin density distributions can be experimentally
observed. Spin-polarized scanning tunneling microscopy\cite{Yamada03}
with magnet-coated tips or optical techniques such as the spatially
resolved Faraday rotation spectroscopy\cite{Kato04} can produce
high-resolution images of spin density profiles.

\section{Conclusion\label{sec:c}}

We have investigated the symmetry properties of eigenstates in
semiconductor quantum wires and observed the charge and spin density
modulations along quantum wires consisting of a
ferromagnet/semiconductor junction by using the coherent scattering
theory.  We have shown that the Rashba or Dresselhaus SO coupling
terms induce charge density oscillations perpendicular to the
propagating direction and that the (anti)symmetric structure of charge
and spin distributions in eigenstates is deeply related to the
oscillations. Charge and spin density oscillations can be
experimentally observed as long as the SO coupling strength or the
voltage drop across the junction are small enough.

Impurities, dirty interfaces, and electron-electron interactions
that are not included in our study
affect the transport in low-dimensional systems.  Although these
effects will therefore influence the form of the charge density
modulations and make it more homogeneous, the charge density
modulation may still be observed in sufficiently clean and low-density
samples.

\begin{acknowledgments}
We would like to thank W. Belzig, M.-S. Choi, and J. Schliemann
for helpful discussions.
This work was financially supported by the SKORE-A program, the Swiss
NSF, and the NCCR Nanoscience.
\end{acknowledgments}

\end{document}